\documentclass[11pt,twoside]{article}

\newcommand{\teff}{T$_{eff}$}

\usepackage{asp2006}
\usepackage{epsf}
\usepackage{psfig}
\usepackage{lscape}

\begin{document}
\title{Ultracool Subdwarfs: \\ Subsolar Metallicity Objects Down to Substellar Masses}   
\author{Adam J.\ Burgasser}   
\affil{Massachusetts Institute of Technology}    

\begin{abstract} 
In the past few years, astronomers have uncovered several very
low-temperature, metal-poor stars with halo or thick disk kinematics and peculiar spectral and photometric properties, so-called
ultracool subdwarfs.  These include the first examples of L subdwarfs - metal-poor analogs of the L dwarf spectral class - and 
slightly metal-deficient T dwarfs.  Ultracool subdwarfs
provide useful empirical tests of low temperature atmosphere and evolutionary models, and are probes of the halo mass function down to and below the (metal-dependent) hydrogen burning limit. 
Here I summarize the optical and near-infrared 
spectroscopic properties of these objects, 
review recent research results, and point out scientific issues
of interest in this developing subject. 
\end{abstract}




\section{Introduction}

Ultracool subdwarfs (UCSDs) are metal-deficient, very low mass stars and brown
dwarfs with late spectral types.  
They are the metal-poor analogs of ultracool
dwarfs (spectral types M7 and later) 
and represent the low effective temperature 
({\teff} $\la$ 3000~K; \citealt{leg00})
extensions of the M subdwarf (sdM) and extreme subdwarf
(esdM) classes \citep{giz97}.
Cool and ultracool subdwarfs typically exhibit 
halo kinematics, 
and were presumably formed early in the
Galaxy's history.  
These low mass objects are
important tracers of Galactic structure and chemical enrichment history 
and are representatives of the first generations of star formation.
UCSDs encompass the new spectral class of L subdwarfs (sdL;
\citealt{me0532,lep1610}) and metal-poor T dwarfs 
(e.g., \citealt{metgrav}), reaching masses
below the
hydrogen burning minimum mass.

This contribution provides an update on the state
of UCSD research since a first review was made at Cool Stars 13 by
\citet{mecs13}.  Here I focus on observed spectral
properties ($\S$2), new discoveries ($\S$3) and considerations
for spectral classification ($\S$4),
and then briefly touch on other issues 
under investigation ($\S$5).

\section{Ultracool Subdwarf Spectral Properties}

Like ultracool dwarfs, UCSDs
exhibit complex optical and near-infrared
spectral energy distributions dominated
by strong, overlapping molecular absorption bands; numerous 
neutral metal line features; and red optical spectral continua
(see~\citealt{giz97,lep03,lep0822,sch1013,sch1444,megmos}).  
They are distinguished spectroscopically by signatures of
metal deficiency, notably enhanced metal hydride and weakened
metal oxide absorption bands \citep{mou76},
and blue near-infrared colors resulting from 
collision induced H$_2$ absorption \citep{sau94}.
These spectral peculiarities are more pronounced in 
lower metallicity subdwarfs
(Figure~1).  The reduction in TiO and VO opacity at optical
wavelengths allows for the emergence of several weaker lines 
not generally seen in solar metallicity M dwarfs, including Ca~I,
Ca~II, Rb~I, and Ti~I.  
The smooth continuum opacity of H$_2$ results
in damped H$_2$O and CO bands
at near-infrared wavelengths, so the latest-type and most
metal-poor UCSDs exhibit
bland spectral energy distributions
longward of 1.5 $\micron$ \citep[see Figure 1]{leg00,cus06}.

\begin{figure}[!ht]
\plotfiddle{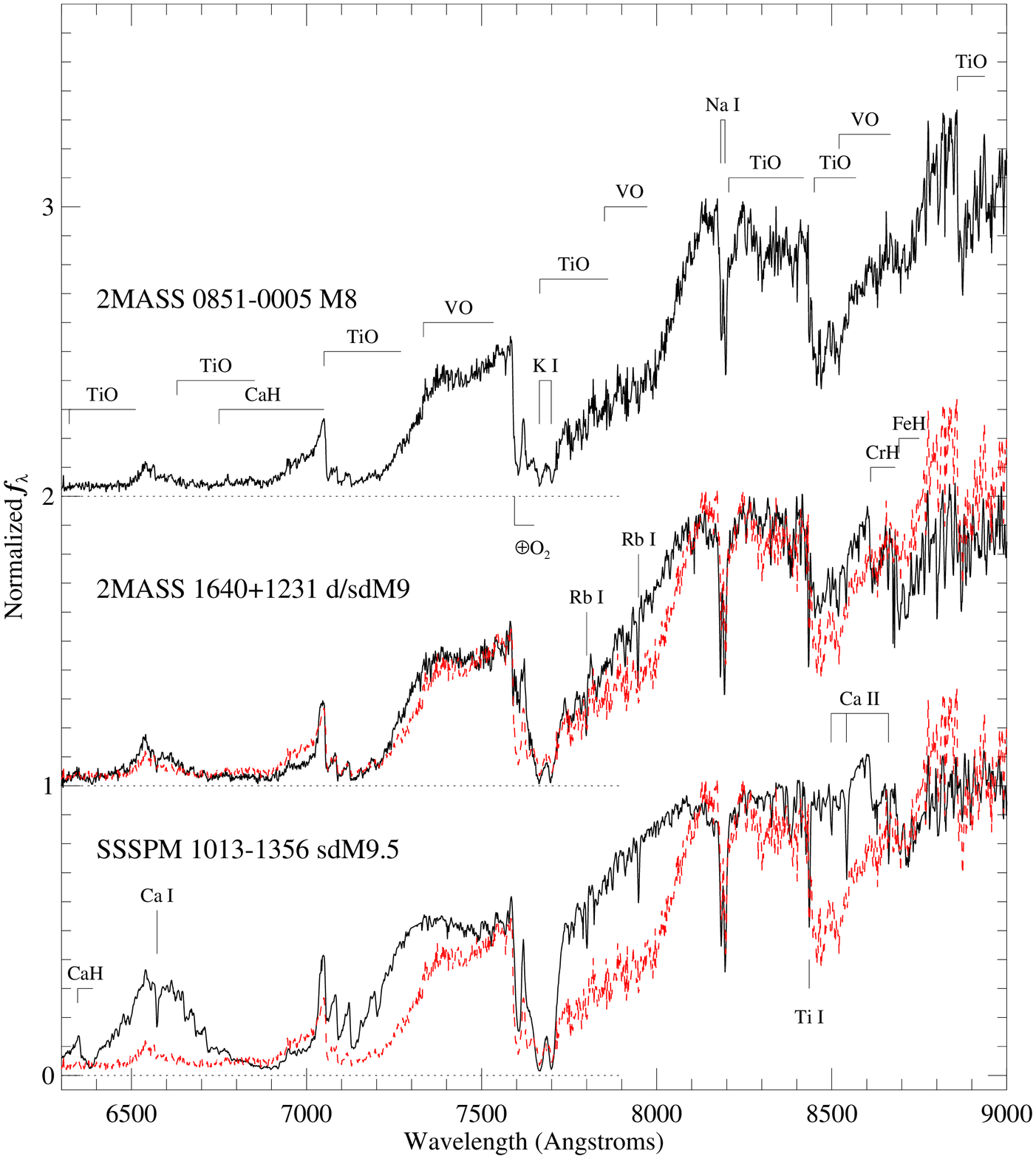}{1in}{0}{25}{25}{-180}{-90}
\plotfiddle{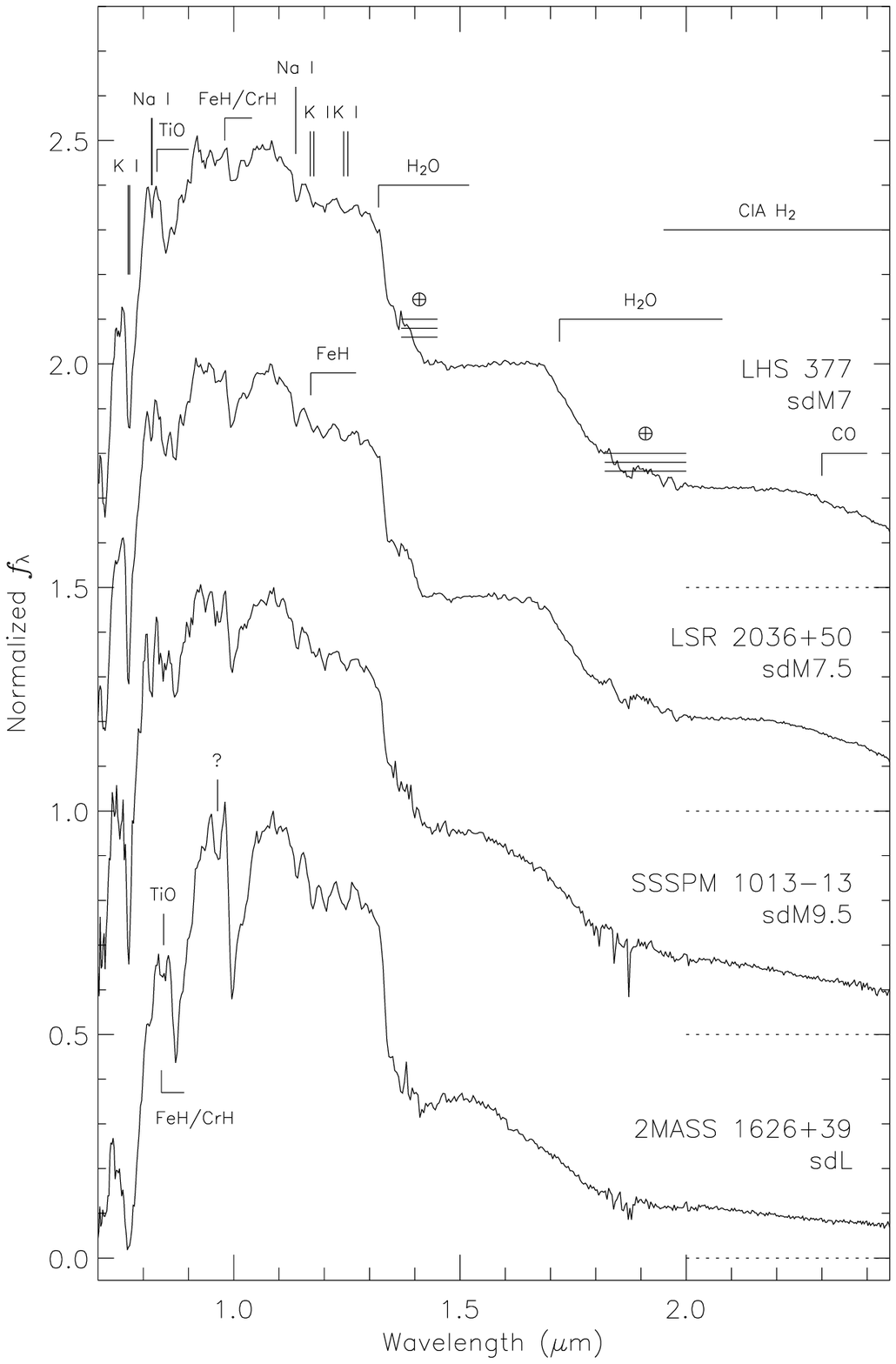}{1in}{0}{33}{33}{10}{-10}
\caption{
(Left) Comparison of M dwarf (top, and dashed lines),
mild subdwarf (middle) and subdwarf (bottom) spectra, illustrating how
reduced metallicity and metal oxide opacity 
allows for the emergence of weaker atomic metal lines
This figure also demonstrates that
intermediate metallicity 
``mild'' (d/sd) UCSDs are readily distinguishable from 
their dwarf and subdwarf counterparts
(from \citealt{megmos}).
(Right) Low resolution near-infrared spectra of four UCSDs.  Note
the relatively smooth and featureless spectra of the two
latest-type objects beyond 1.5~$\micron$,
caused by strong H$_2$ absorption, and the peak in flux
around 1~$\micron$
(from \citealt{me1626}).
} 
\end{figure}

\citet{rei06} have recently presented high resolution spectra for
UCSDs, providing the first rotational velocity
and accurate radial velocity measurements for these sources.  The
high $v\sin{i}$ = 65$\pm$15~km/s of the L subdwarf 2MASS~J0532+8246 
\citep{me0532} suggests that the lowest mass halo stars may not lose
their angular momentum as efficiently as more massive stars.
They also find that 
one of the first L subdwarfs to be identified, LSR~1610-0040 
\citep{lep1610}, has mixed M dwarf and subdwarf
spectral features, a conclusion also reached by \citet{cus06} on the 
basis of near-infrared spectroscopy.  This source appears to 
be peculiar, even among the small number of UCSDs now known.

\section{New Discoveries}

\citet{megmos} provides a compendium of 16 currently known UCSDs,
an increase of 60\% since
Cool Stars 13.  This increase has been driven by 
new proper motion search programs using 
red optical photographic plate surveys
(e.g., SUPERBLINK, \citealt{lep02}; and
the SuperCOSMOS Sky Survey [SSS], \citealt{ham01})
and serendipitous discoveries in near-infrared catalogs such
as 2MASS.
New discoveries include two ultracool
extreme subdwarfs \citep{melehpm,megmos}, 
tripling the number of known ultracool esdMs;
and a new L subdwarf,
SDSS 1256-0224, reported by \citet{siv07}.

While red optical proper motion surveys have been successful, 
the spectral
energy distributions of cooler UCSDs peak around 1~$\micron$
(a balance of reduced {\teff} and increased absorption by H$_2$;
see Figure~1),
so near-infrared measurements become increasingly useful.  Indeed,
the UKIDSS survey with its 1~$\micron$ $Y$-band should be 
highly sensitive to UCSDs \citep{hew06}.  
The recent inception of near-infrared proper motion surveys
(e.g.~\citet{dea05,art06}; J.\ D.\ Kirkpatrick, 2007, in prep.)
will likely expand the known UCSD population substantially
in the near term.

\section{Ultracool Subdwarf Classification}

Spectral types for UCSDs are currently based on extrapolations
of schemes developed for earlier-type M subdwarfs \citep{giz97}, 
based on TiO and CaH bands in the 6200--7300~{\AA} region.  
These features become inadequate in the UCSD regime
as TiO and CaH bands decline due to condensation 
(e.g.~\citealt{lod02}); 
wide, pressure-broadened
{Na~I} and {K~I} lines suppress adjacent features \citep{bur03};
and the sources themselves become exceedingly faint at optical wavelengths.
\citet{giz06} and 
\citet{megmos} have suggested the use of the same red optical features
used to classify L dwarfs as diagnostics for UCSD classification, and 
the direct comparison of L subdwarf spectra to 
L dwarf spectral standards (see Figure~2).
At near-infrared
wavelengths, useful diagnostics are largely limited to the $J$-band
region (Figure~1), and work is progressing in this direction
(Burgasser et al., in prep.).

\begin{figure}[!ht]
\plotfiddle{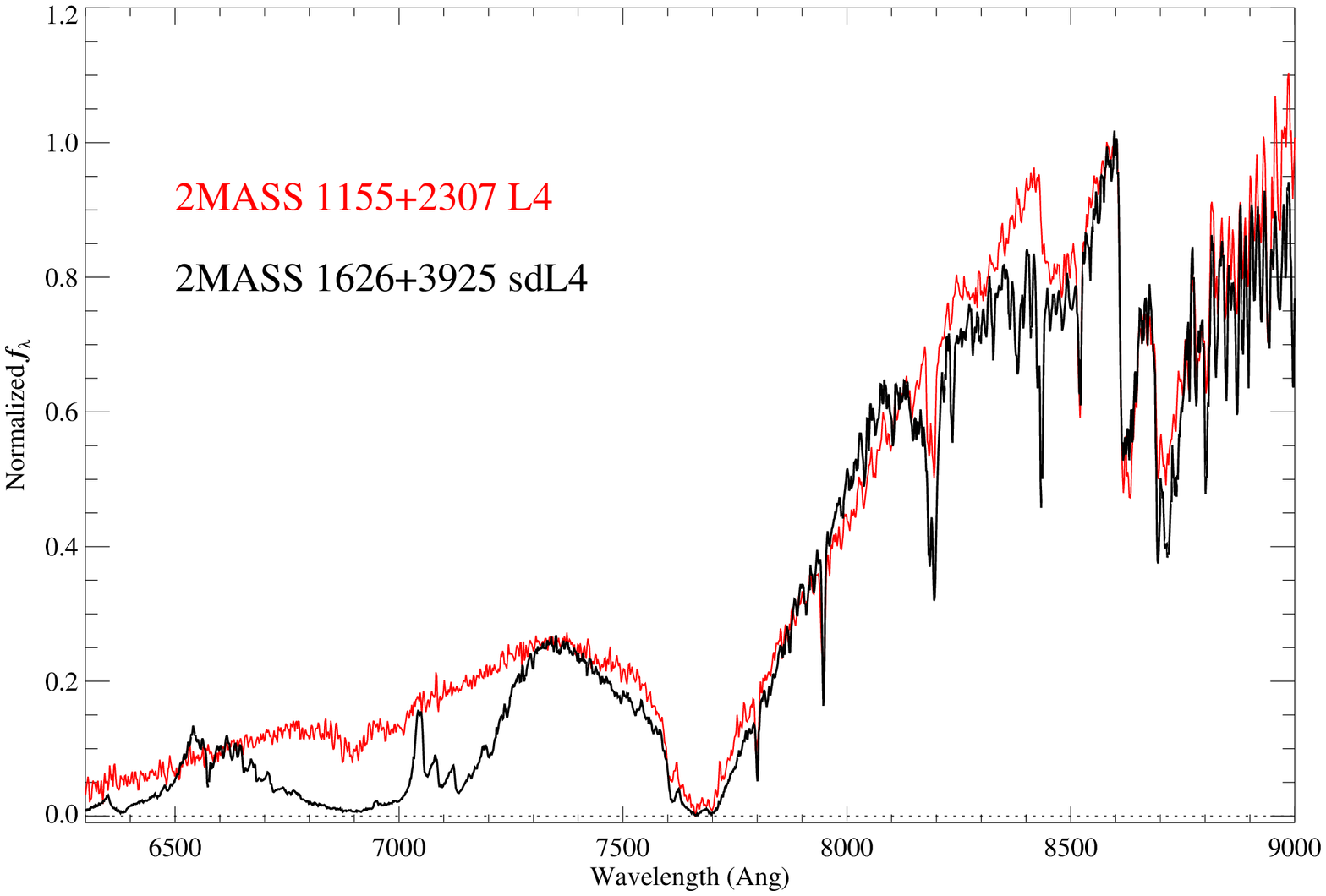}{0.6in}{0}{27}{27}{-180}{-60}
\plotfiddle{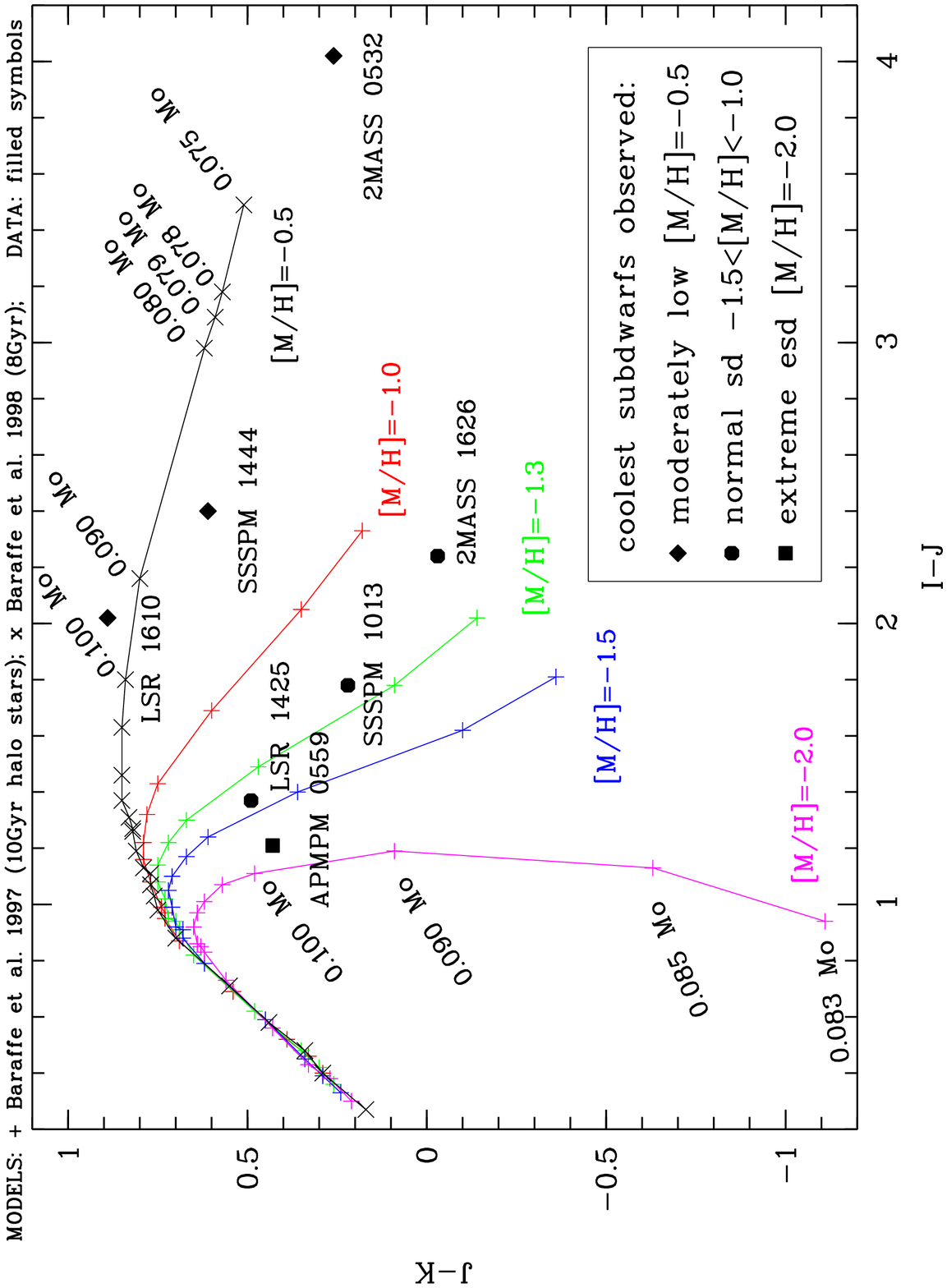}{0.6in}{-90}{23}{23}{-10}{120}
\caption{
(Left) Red optical spectra of the L subdwarf 2MASS~J1626+3925 (black line)
and the L4 dwarf spectral standard 2MASS~J1155+2307
(red dashed line). 
Their rough agreement
suggests a means of classifying 
L subdwarfs \citep{giz06,megmos}, with deviations indicative
of metallicity effects  
(adapted from \citealt{megmos}).
(Right) Color-color diagram of known UCSDs compared to metal-dependent
evolutionary and atmospheric models.  While current
models may not accurately match observations ($\S$5), 
it is interesting to note the range
in metallicities suggested by this comparison (from \citealt{sch1444}).
}
\end{figure}

An issue to consider in the classification of UCSDs is 
the division of metallicity classes.
Strong, blanketing molecular bands
results in greater spectral sensitivity to metallicity and chemistry effects.
The current population of UCSDs may already span a fairly broad range of
metallicities, as suggested by comparisons to atmospheric models
\citep[see Figure~2 above]{sch1444}; and the metallicities of early-type
sdMs may
not (and perhaps need not) correspond to those of late-type sdMs and
sdLs.
In addition, even slightly metal-poor ([M/H] $\ge$ -0.5) 
``mild'' UCSDs
are spectrally distinct (Figure~1), and a ``d/sd'' designation for these
objects has been suggested. This group would encompass slightly
metal-poor T dwarfs, 
distinguished by depressed $K$-band peaks (due
to H$_2$ absorption) and enhanced $Y$-band peaks (possibly related to
the red wing of the K~I doublet at 0.77~$\micron$; 
\citealt{metgrav}); and 
several recent discoveries of so-called ``blue'' L dwarfs 
\citep{cru03,kna04,chi06} that may also be metal-poor.


\section{Other Issues}

I briefly summarize other issues that have arisen in recent
studies
of UCSDs.  The interested reader is encouraged to examine the literature
for further discussion.

\smallskip \noindent {\em Dust Formation:}
One of the defining properties of L field dwarfs is the formation
of condensate dust in their photospheres, driving the depletion
of gaseous TiO, VO, CaH, Al~I, Ca~I and Ti~I in the photosphere
\citep{bur99,lod02}.  However, several late-type subdwarfs
exhibit features from these species when it is absent in their
dwarf counterparts 
\citep[see Figure~2 above]{me0532,sch1444,rei06}.
Their persistence suggests that
condensate formation may be inhibited in metal-poor atmospheres;
chemical modeling
of this effect is needed. 


\smallskip \noindent {\em Spectral Modeling:} Attempts
to derive physical properties for UCSDs based on 
spectral model fits have met with mixed results 
(e.g.~\citealt{sch99,lep0822,megmos}).  
The difficulties stem from the same issues that arise when 
modeling ultracool dwarf spectra; e.g.,
complex molecular opacities, and
chemistry and condensate grain formation.
There are new concerted
efforts to develop better metal-poor atmosphere models in the UCSD regime
(\citealt{bur06}; Schweitzer, Hauschildt \& Wawrzyn, these proceedings).

\smallskip \noindent {\em Binaries:} Binary systems are useful
laboratories for studying detailed properties, testing
formation theories and finding low temeprature
sources as companions (see splinter session overview by H.\ Bouy).
There have been a couple of studies targeting M subdwarfs for companions
\citep[Gelino \& Kirkpatrick, these proceedings]{zap04}
with as yet no reported UCSD discoveries.

\smallskip \noindent {\em Distance Measurements:} These are perhaps
the most crucial observations needed, as only two UCSDs have
reported parallaxes.
Such measurements 
make possible the construction of absolute magnitude, luminosity
and {\teff} scales, and facilitate comparisons to spectral and structural 
models.

\acknowledgements 

I would like to thank A.\ Burrows, J.\ D.\ Kirkpatrick,
R.-D.\ Scholz, and A.\ Schweitzer for discussions in the final preparation
of this contribution, and R.-D.\ Scholz 
for providing an electronic version of Figure~2. 



\end{document}